\documentclass[lettersize,journal]{IEEEtran}
\usepackage{amsmath,amsfonts}
\usepackage{algorithmic}
\usepackage{algorithm}
\usepackage{array}
\usepackage{subcaption}
\usepackage{textcomp}
\usepackage{stfloats}
\usepackage{url}
\usepackage{verbatim}
\usepackage{graphicx}
\usepackage{cite}
\usepackage{booktabs}
\usepackage{multirow}
\usepackage{colortbl}
\usepackage[table]{xcolor}
\usepackage{hyperref} 
\begin{document}

\title{Redefining Digital Twins as Predictive Decision Engines for AI-Native Wireless Networks}

\author{Afan Ali,~\IEEEmembership{Member,~IEEE}, Ali Arshad Nasir,~\IEEEmembership{Senior Member,~IEEE}, Naveed Iqbal,~\IEEEmembership{Senior Member,~IEEE}, \\ and Daniel Benevides da Costa,~\IEEEmembership{Senior Member,~IEEE}
\thanks{The authors are with the Interdisciplinary Research Center for Communication Systems and Sensing (IRC-CSS), Department of Electrical Engineering, King Fahd University of Petroleum and Minerals (KFUPM), Dhahran 31261, Saudi Arabia (e-mail: afan.ali@kfupm.edu.sa; anasir@kfupm.edu.sa;\\ naveediqbal@kfupm.edu.sa; danielbcosta@ieee.org).}}

\markboth{Journal of \LaTeX\ Class Files,~Vol.~, No.~, May~2026}%
{Ali \MakeLowercase{\textit{et al.}}: Redefining Digital Twins as Predictive Decision Engines for AI-Native Wireless Networks}

\IEEEpubid{}

\maketitle

\begin{abstract}
Future artificial intelligence (AI)-native 6G networks require wireless systems that move beyond reactive optimization toward autonomous, predictive, and continuously adaptive intelligence. Most existing digital twin (DT) frameworks use AI only to improve modeling, data generation, or isolated optimization tasks, leaving the DT itself in a passive, synchronization-only role. This article redefines the DT as a predictive decision engine, which is model-agnostic framework that fuses continuous synchronization, predictive reasoning, autonomous decision-making, and closed-loop wireless control into a single system. Reasoning over synchronized network state, the framework anticipates future conditions and acts autonomously before performance degrades, using generative adversarial networks (GANs), large language models (LLMs), diffusion models, or other learning-based engines interchangeably as the underlying predictor. An illustrative unmanned aerial vehicle (UAV)-assisted non-terrestrial network (NTN) deployment, using a lightweight conditional generative adversarial network (cGAN) as one illustrative predictor, demonstrates the practical effectiveness of the proposed framework by achieving considerable energy savings over reactive baselines while maintaining reliable quality of service (QoS) under highly dynamic conditions.
\end{abstract}

\section{Introduction}

\IEEEPARstart{S}{ixth}-generation (6G) wireless networks are evolving into distributed cyber--physical ecosystems that tightly integrate communication, sensing, computing, localization, and intelligent decision-making. Emerging applications, including integrated sensing and communication (ISAC), extended reality (XR), autonomous transportation, unmanned aerial vehicle (UAV)-assisted non-terrestrial networks (NTNs), and space--air--ground--sea (SAGS) systems, generate massive heterogeneous data under highly dynamic wireless conditions~\cite{ansari_pioneering_2025}. Unlike earlier generations, these networks must continuously adapt to mobility, traffic, interference, and energy constraints, making autonomous, system-level intelligence a requirement rather than an option.

Addressing these challenges, the convergence of generative artificial intelligence (GenAI) and digital twins (DTs) has emerged as a promising direction for intelligent wireless networks. DTs offer continuously synchronized virtual representations of physical systems, while GenAI enables prediction, reasoning, and knowledge generation from large-scale network data~\cite{khan_digital-twin-enabled_2022,nguyen_digital_2021}, expanding the DT's role from passive visualization toward learning, simulation, and optimization. Industry is moving in the same direction, as shown in NVIDIA's Omniverse and AI Aerial platforms, which target large-scale wireless emulation. Similarly, Keysight has introduced radio-frequency (RF) DT platforms for AI-assisted testing, and the 3rd Generation Partnership Project (3GPP) has begun standardizing Network Digital Twins (NDTs) for AI-native networks~\cite{cohenarazi2025nvidiaaiaerialainative,keysight2026rfdt,3gpp28915}. Table~\ref{tab:related} summarizes this evolution, in which early work used GenAI mainly to improve DT fidelity through synthetic data and prediction~\cite{naeem_survey_2025,savaglio_generative_2025}, while more recent studies extend the synergy toward intelligent wireless DTs that support network management, orchestration, and resource optimization~\cite{guan_integrating_2025,singh_wind_2025,cao_decentralized_2026,zhao_multi-layer_2026}.

Despite this progress, the evolution remains incomplete. Existing GenAI--DT frameworks mainly use AI to improve DT fidelity, prediction accuracy, or network optimization, so the DT still serves as a synchronized replica or an optimization aid rather than a decision-maker. Prediction alone is not autonomous intelligence. What remains unexplored is redefining the DT itself into a predictive decision engine that anticipates future network conditions, reasons over their evolution, and proactively adapts wireless resources through closed-loop feedback before degradation occurs. This capability is increasingly critical for dynamic environments such as UAV-assisted NTNs, ISAC, and AI-native 6G, where resource decisions must often be made ahead of rapidly changing conditions, and is consistent with the 3GPP Release~18 and Release~19 roadmap on AI-native management, NDTs, and Closed Control Loops (CCLs) discussed further in Section~\ref{sec:framework}.

Motivated by this gap, this article redefines the role of the digital twin itself: rather than a synchronized replica or an optimization aid, the DT becomes a predictive decision engine at the center of a unified, closed-loop, AI-native framework. The proposed framework integrates continuous DT synchronization, predictive reasoning, autonomous decision-making, and proactive wireless resource orchestration into one system that is intentionally agnostic to the underlying learning model, accommodating GANs, LLMs, diffusion models, foundation models, or future autonomous AI agents interchangeably as wireless intelligence evolves. This redefinition also aligns with the trajectory of 3GPP standardization, where NDTs introduced under Release~19 and extended in Release~20, are framed as management-plane capabilities that the proposed synchronization, prediction, and decision layers are well positioned to realize~\cite{3gpp28915}. As one concrete instantiation, a lightweight conditional generative adversarial network (cGAN) illustrates the framework in a UAV-assisted NTN use case, showing how future network states can be inferred to proactively optimize transmission power and energy-efficient resource allocation, the underlying contribution, however, is the architectural redefinition of the DT as a decision-maker, not the choice of predictor.

\begin{table*}[!t]
\caption{Evolution of the Synergy Between Generative AI and Digital Twins for Intelligent Wireless Networks.}
\label{tab:related}
\renewcommand{\arraystretch}{1}
\setlength{\tabcolsep}{4pt}
\definecolor{dtrow}{RGB}{235,244,255}
\definecolor{sysrow}{RGB}{255,248,230}
\definecolor{proprow}{RGB}{228,244,228}
\definecolor{headercolor}{RGB}{30,80,160}
\centering
\scriptsize
\begin{tabular*}{\textwidth}{@{\extracolsep{\fill}}
>{\raggedright\arraybackslash}p{6.3cm}
>{\raggedright\arraybackslash}p{11.4cm}
@{}}
\toprule
\rowcolor{headercolor}
\textcolor{white}{\textbf{Evolution Stage and Representative Work}} &
\textcolor{white}{\textbf{Role of the GenAI--DT Synergy}} \\
\midrule
\rowcolor{dtrow}
GenAI-Enhanced Digital Twins~\cite{guan_integrating_2025,savaglio_generative_2025,nguyen_digital_2021} &
The DT acts as a synchronized digital replica. GenAI operates mainly at the signal level, improving fidelity through synthetic data generation and channel or traffic prediction, while all control decisions remain external to the twin. \\
\midrule
\rowcolor{sysrow}
AI-Driven Wireless Digital Twins~\cite{singh_wind_2025,cakir_d-ndt_2026,cao_decentralized_2026,lin_ai-native_2025} &
The DT becomes an intelligent optimization platform. GenAI supports network management, orchestration, and resource optimization at the system level, but prediction and decision-making remain loosely coupled, separate processes. \\
\midrule
\rowcolor{proprow}
\textbf{Predictive Closed-Loop DT (This Work)} &
\textbf{The DT is redefined as a predictive decision engine. Continuous synchronization, predictive reasoning, autonomous decision-making, and closed-loop wireless control are fused into one model-agnostic framework, so the twin itself anticipates and acts rather than merely informing external controllers.} \\
\bottomrule
\end{tabular*}
\end{table*}

\section{Evolution of the Synergy Between Generative AI and Digital Twins}
\label{sec:evolution}
This section traces the GenAI--DT synergy through three evolutionary stages, summarized in Table~\ref{tab:related} and Fig.~\ref{fig:overall}.

\subsection{From Synchronized Digital Twins to Intelligent Network Awareness}
\begin{figure*}[t]
\centering
\captionsetup[subfigure]{
    font={footnotesize},
    labelfont=bf,
    justification=centering,
    skip=4pt
}

\begin{subfigure}[b]{0.48\textwidth}
    \centering
    \fbox{\includegraphics[width=0.96\linewidth,
                           height=6.2cm,
                           keepaspectratio]{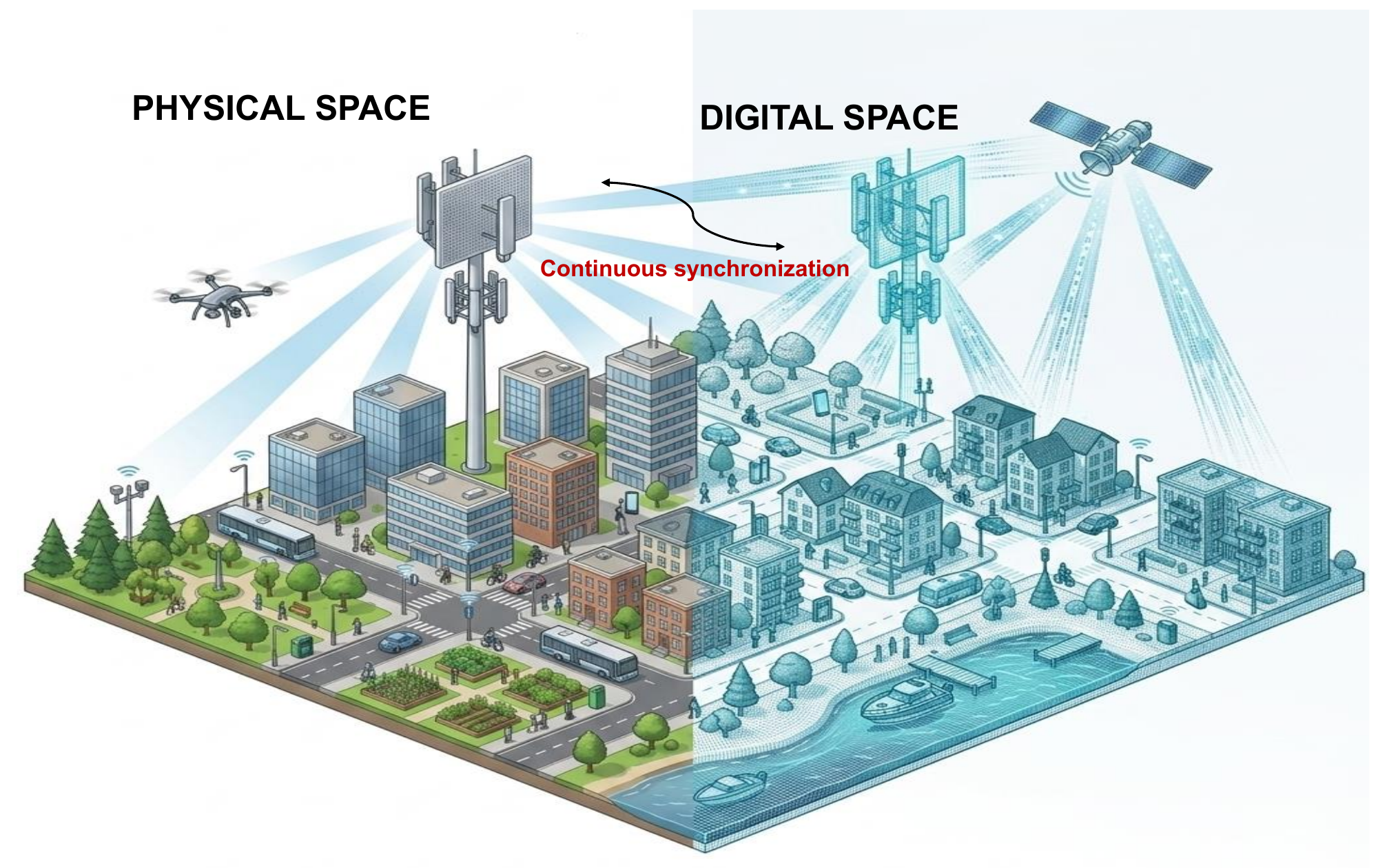}}
    \caption{Conventional digital twin.}
    \label{fig:1a}
\end{subfigure}
\hfill
\begin{subfigure}[b]{0.48\textwidth}
    \centering
    \fbox{\includegraphics[width=0.96\linewidth,
                           height=6.2cm,
                           keepaspectratio]{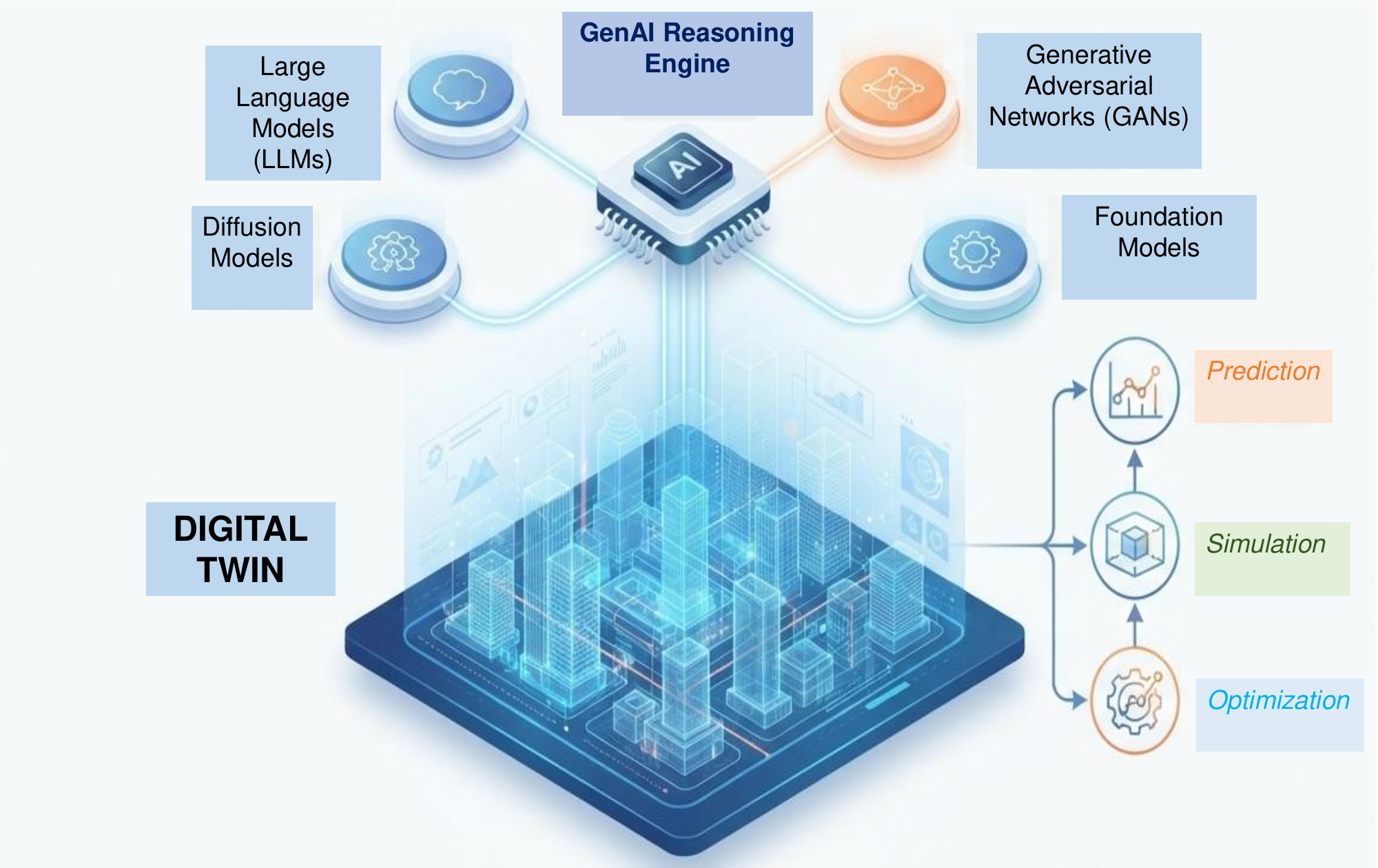}}
    \caption{GenAI-enhanced digital twin.}
    \label{fig:1b}
\end{subfigure}

\vspace{4mm}

\begin{subfigure}[b]{0.6\textwidth}
    \centering
    \fbox{\includegraphics[width=0.96\linewidth,
                           height=6.6cm,
                           keepaspectratio]{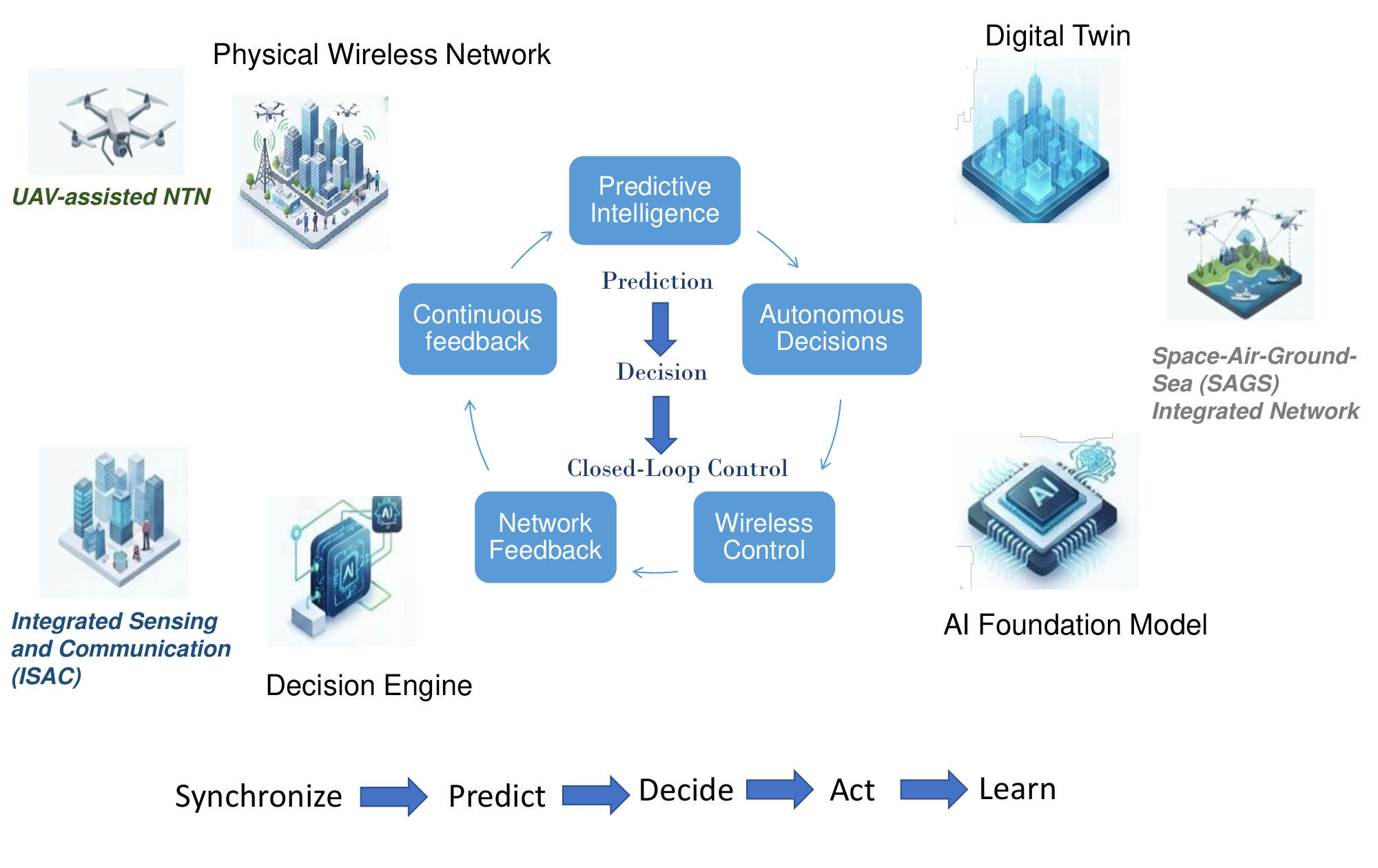}}
    \caption{Predictive closed-loop wireless intelligence.}
    \label{fig:1c}
\end{subfigure}

\vspace{3mm}

\caption{Evolution of the synergy between Generative AI and Digital Twins.
(a)~Stage I: Conventional Digital Twins.
(b)~Stage II: Synergistic GenAI-DT.
(c)~Stage III: Proposed Predictive Closed-Loop Wireless Intelligence.}

\label{fig:overall}

\end{figure*}

The GenAI--DT partnership began with a simple goal, i.e., wireless DTs that can keep pace with the physical world. As shown in Stage~I of Fig.~\ref{fig:1a}, the twin continuously ingests live measurements, including channel state information (CSI), user mobility, traffic demand, environmental sensing, topology, and resource utilization, to maintain a running picture of the network. Unlike static simulation platforms, it gives operators real-time visibility, scenario testing, and what-if analysis without disturbing live service. This is increasingly vital as 6G integrates extremely large-scale multiple-input multiple-output (XL-MIMO), ISAC, reconfigurable intelligent surfaces (RIS), UAV-assisted links, and NTNs, all evolving too fast for offline models to capture. Modern DTs therefore combine 3D environment modeling, propagation awareness, and persistent synchronization, reflected in industry efforts such as NVIDIA's AI Aerial and Omniverse platforms, Keysight's RF DTs, and 3GPP's recognition of NDTs as a key enabler for AI-native 6G.
A synchronized replica has real operational value: operators can read network behavior accurately, trial resource strategies safely, and feed AI models with consistent data. However, Stage~I twin remains passive. Decisions on resource allocation, power control, and mobility management are still delegated to external optimizers that treat the DT as a data source rather than a decision partner. As networks grow more dynamic, this separation between awareness and action becomes a limitation that motivates the next phase of the GenAI--DT evolution.

\subsection{The Rise of the GenAI--DT Synergy}

GenAI pushed wireless DTs into a new phase. Rather than simply mirroring the physical network, researchers began embedding generative models directly into the DT to help it learn, predict, and optimize. As Stage~II of Fig.~\ref{fig:1b} illustrates, GenAI moves the twin beyond synchronized awareness into prediction, synthetic data generation, and intelligent optimization, fueled by progress in foundation models, LLMs, diffusion models, and GANs that capture the statistical behavior of wireless systems. The DT can now synthesize realistic network states, fill in missing information, and forecast future conditions, growing from a signal-reconstruction aid into a core intelligence engine.
Early GenAI--DT frameworks used generative models mainly to boost twin fidelity through synthetic data and channel prediction~\cite{guan_integrating_2025,savaglio_generative_2025}, while later studies built intelligent wireless DTs supporting network orchestration, distributed optimization, and AI-assisted resource management~\cite{singh_wind_2025,cao_decentralized_2026,zhao_multi-layer_2026}. The DT is becoming an optimization platform rather than a passive replica, but the mission is unchanged: sharper predictions and better optimization, handed off to separate controllers that decide what the network actually does. The twin has grown smarter, but it still advises rather than acts.

\subsection{Toward Predictive Closed-Loop Wireless Intelligence}
The progress of GenAI-enhanced DTs raises a natural question: what comes after intelligent optimization? Existing frameworks have made real strides in twin fidelity, prediction accuracy, and optimization, but prediction is only half the story. In most setups, the predicted state is still handed off to external optimizers or management entities that make the actual control decisions, so synchronization, prediction, optimization, and actuation remain separate steps.
This separation grows more problematic as networks move toward AI-native 6G. Scenarios such as UAV-assisted NTNs, ISAC, and dense heterogeneous deployments face channel conditions, mobility, traffic, and resource availability that shift continuously, and reacting to the present state is often too slow. The network needs to anticipate what is coming and act before performance slips, which requires the DT to advance beyond synchronized awareness into predictive decision-making within a continuous closed loop.
Stage~III of Fig.~\ref{fig:1c} captures this transition. Where earlier stages used GenAI mainly to improve the twin itself, the proposed framework redefines the DT as the decision-maker at the heart of the control loop: continuous synchronization keeps the virtual and physical networks aligned, predictive reasoning anticipates future conditions from real-time and historical observations, and those forecasts convert directly into actions such as power adaptation, beam management, mobility support, spectrum allocation, and energy-aware orchestration. Feedback from the physical network flows continuously back into the twin, so predictions and decisions evolve alongside the network itself.
This goes beyond better predictions as it redefines what the DT does. Rather than a synchronized replica or an optimization platform, the DT becomes a predictive decision engine that senses, predicts, decides, acts, and learns from network behavior in a continuous cycle, providing the conceptual foundation for the unified framework presented in Section~\ref{sec:framework}.

\begin{figure*}[t]
\centering
\captionsetup[subfigure]{
    font={footnotesize},
    labelfont=bf,
    justification=centering,
    skip=4pt
}

\begin{subfigure}[b]{0.33\textwidth}
    \centering
    \fbox{\includegraphics[width=1.1\linewidth,
                           height=13cm,
                           keepaspectratio]{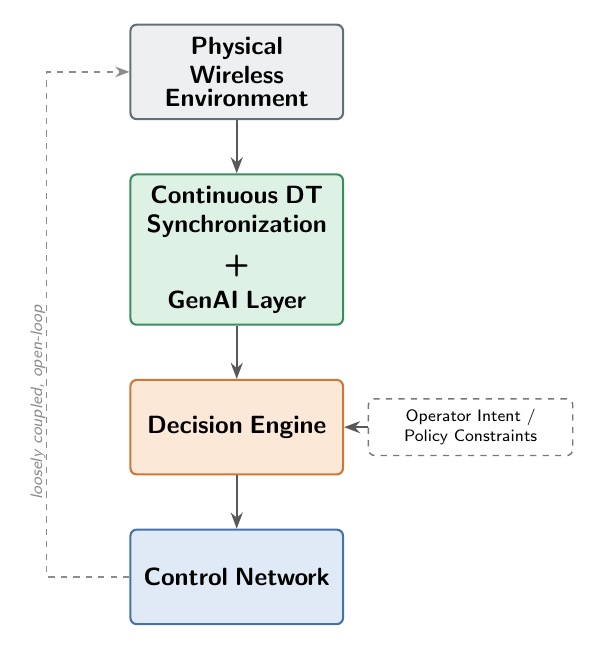}}
    \caption{Conventional GenAI--DT pipeline.}
    \label{fig:2a}
\end{subfigure}
\hfill
\begin{subfigure}[b]{0.62\textwidth}
    \centering
    \fbox{\includegraphics[width=1.1\linewidth,
                           height=7.6cm,
                           keepaspectratio]{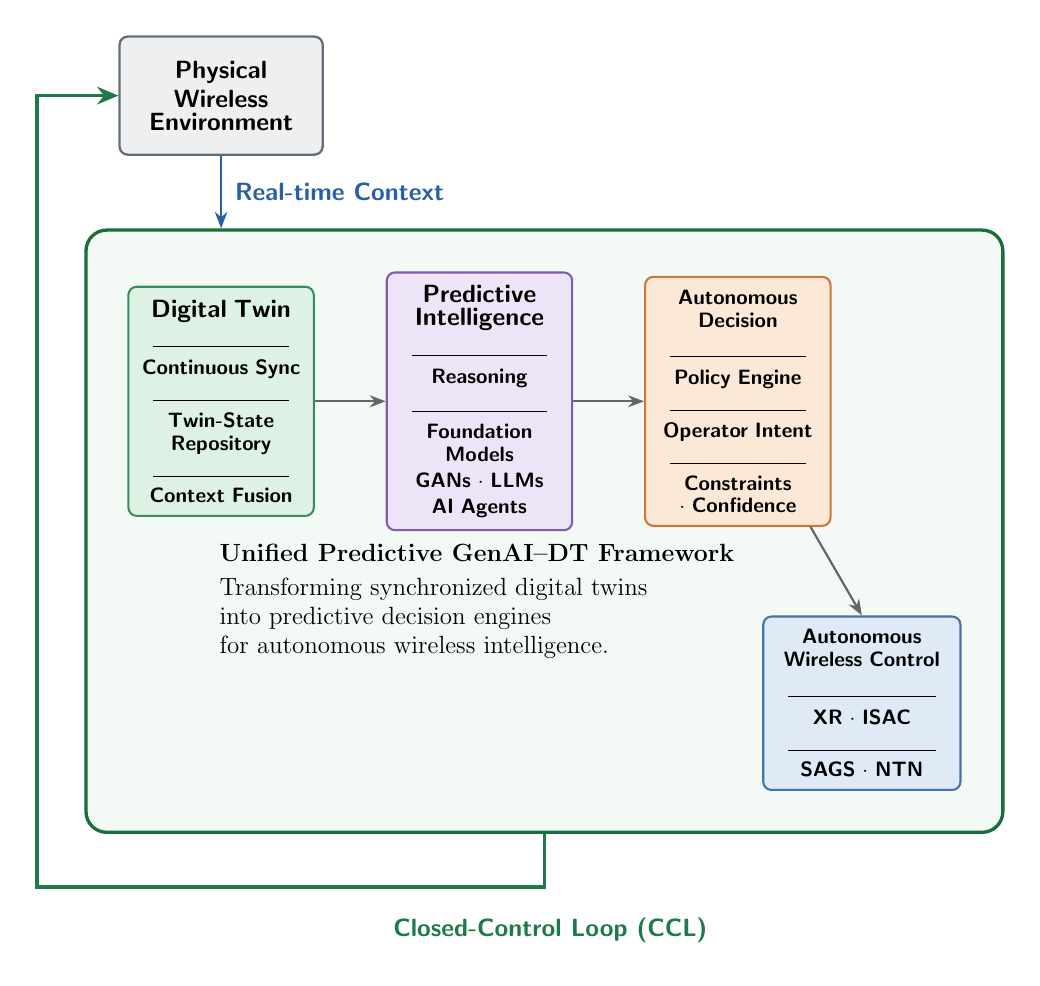}}
    \caption{Proposed unified predictive decision engine.}
    \label{fig:2b}
\end{subfigure}

\vspace{3mm}

\caption{From conventional GenAI-assisted digital twins to the proposed predictive decision engine.
(a)~Conventional pipeline, where the digital twin feeds synchronized state and predictions to separate optimizers.
(b)~Proposed framework, in which synchronization, prediction, decision-making, and control operate as one continuously learning system.}

\label{fig:framework}

\end{figure*}

\section{A Unified Predictive Decision Engine}
\label{sec:framework}

Building on the evolution outlined above, we now present the architectural realization of the Stage~III paradigm. The core contribution here is architectural rather than model-specific. Fig.~\ref{fig:2a} summarizes the conventional GenAI--DT pipeline used in prior work, where prediction and control remain separate steps and Fig.~\ref{fig:2b} shows the proposed redefinition of the DT itself as a predictive decision engine, in which continuous synchronization, predictive intelligence, autonomous decision-making, and closed-loop wireless control operate within a single, model-agnostic framework. 

\subsection{Continuous GenAI--DT Synchronization}
The framework rests on continuous GenAI--DT synchronization, with one critical difference from existing approaches: synchronization does more than keep the twin accurate for prediction or optimization. As Fig.~\ref{fig:2b} shows, each cycle both updates the twin state and feeds contextual information forward to the predictive intelligence layer, so future network evolution is inferred before any control decision is made.
The synchronized state fuses heterogeneous observations from the physical network, including CSI, user mobility, traffic demand, propagation conditions, environmental sensing, and resource utilization, into a unified system-level picture of how communication, sensing, computing, mobility, and energy interact and evolve.
A defining feature is the bidirectional link between the twin and the predictive engine: the twin supplies fresh observations to the predictor, while predicted future states fold back into the evolving DT representation and guide subsequent decisions. The twin is no longer a synchronized replica; it becomes an active intelligence platform bridging observations, prediction, and control.
To learn robustly across varied conditions, the predictive engine trains on a hybrid dataset combining high-fidelity ray-tracing-based channel and mobility realizations, generated with tools such as Sionna RT under the 3GPP TR~38.821 propagation assumptions~\cite{3gpp38821}, with DT-generated synthetic samples that extend coverage to rare events, such as sudden congestion, sharp mobility transitions, and severe interference, that practical measurement campaigns seldom capture. The ray-traced component serves as a higher-fidelity reference against which the DT's own propagation representation is checked, rather than being mixed indiscriminately into training, which keeps the evaluation of the twin's fidelity independent of the twin's own synthetic-generation process and avoids validating a model against data it helped produce. Fully validating DT fidelity against physical field measurements remains important future work, the present study validates the closed-loop control logic under the twin's own propagation model rather than the propagation model itself.
Practical deployments face communication overhead, synchronization latency, and transient mismatches between physical and virtual domains. Rather than assuming perfect lockstep, the framework works with incrementally refreshed twin states that evolve as measurements arrive, keeping predictive reasoning effective despite delays while reducing the burden of full-state updates after every event.

The proposed synchronization layer also maps naturally onto ongoing 3GPP standardization of NDTs. Release~19 introduced NDTs through study on management aspects of NDTs~\cite{3gpp28915}, with a Release~20 phase-2 study now extending this work. These efforts frame the NDT as a management-plane capability, and the synchronization, prediction, and decision layers proposed here are naturally positioned as candidate mechanisms operating within that management-plane scope, offering a concrete pathway for realizing standardized NDTs as autonomous, predictive components of future radio access networks.

\subsection{Predictive Intelligence and Autonomous Decision Making}
With the synchronized twin state in place, the framework moves into predictive intelligence. Rather than reacting to snapshots of network conditions, this layer continuously reasons over the evolving twin state to estimate future communication conditions, traffic, interference, mobility, and energy demand, forming the basis for proactive resource orchestration that lets the network adapt before performance slips.
Unlike conventional predictors that sit apart from network control, this layer is woven into the DT synchronization process: each synchronization update sharpens the twin state, and each prediction immediately shapes subsequent decisions.
The framework is deliberately model-agnostic, in which, any generative model capable of learning complex wireless dynamics, such as GANs, diffusion models, foundation models, multimodal LLMs, or future autonomous AI agents, can serve as the predictive engine without altering the broader synchronization and control framework. This is the central point of the proposed redefinition: the DT's role as decision-maker does not depend on which model instantiates its predictive reasoning.
For concreteness, this article adopts a lightweight conditional GAN (cGAN) as one illustrative predictive engine, chosen for its low inference latency, stable conditional generation, and computational efficiency in real-time wireless management. More demanding models, such as diffusion models or large multimodal foundation models, may offer richer reasoning, but their higher inference complexity currently limits deployment in latency-sensitive control, quantified in Section~\ref{sec:deployment}.
The predictive engine trains offline on the hybrid dataset described above, then deploys for real-time online inference, keeping latency low while learning from both measured and synthetic behavior before going live. Confidence-aware prediction flags low-confidence or out-of-distribution conditions, prompting model updates, operator intervention, or conservative control policies whenever reliability fades.

\subsection{Closed-Loop Wireless Control and Continuous Learning}

The final stage translates predictive intelligence into autonomous wireless control. Unlike existing GenAI--DT frameworks, where prediction and optimization run as separate processes, the proposed framework tightly integrates synchronization, prediction, decision-making, and control within a continuous feedback loop, as illustrated in Fig.~\ref{fig:2b}: every prediction from the intelligence layer immediately feeds the decision engine, letting the network proactively adapt before performance degrades.

Based on the predicted network evolution, the decision engine continuously determines control actions, including transmission power adaptation, beam management, spectrum allocation, user association, mobility support, sleep-mode activation, and energy-aware resource orchestration. Unlike reactive strategies that respond only after congestion, interference, or energy depletion becomes observable, the proposed framework anticipates future conditions and acts before they materialize.

Prediction, decision-making, and control remain continuously coupled throughout operation: every control action influences the physical network, whose updated state is synchronized back into the DT, and the refreshed twin-state feeds the next prediction cycle. This self-evolving process, where synchronization, prediction, decision-making, and learning operate as one system rather than independent modules, distinguishes the proposed framework from previous methodology where predictive models and controllers remain loosely connected.

The framework also supports practical deployment: operator-defined intent policies, network constraints, and quality-of-service (QoS) requirements can be incorporated within the decision layer so autonomous control stays aligned with deployment objectives. Confidence-aware predictions let uncertain conditions trigger conservative control policies, model refresh, or operator intervention whenever reliability falls below acceptable levels.

Although the case study in this article focuses on UAV-assisted NTNs, the framework is application-independent because the decision engine operates on synchronized network-state representations rather than application-specific signal features, extending naturally to ISAC, cell-free massive MIMO, semantic communications, and integrated terrestrial--non-terrestrial networks, as Section~\ref{sec:deployment} discusses further.

Continuous feedback also lets the framework evolve with the physical network over time. Synchronized observations collected during operation can periodically refine the predictive engine, adapting to slowly varying propagation, traffic, mobility, and deployment changes while mitigating long-term model drift, which keeps predictive intelligence consistent with the evolving environment.

\begin{figure*}[t]
\centering
\captionsetup[subfigure]{
    font={footnotesize},
    labelfont=bf,
    justification=centering,
    skip=4pt
}

\begin{subfigure}[b]{0.48\textwidth}
    \centering
    \fbox{\includegraphics[width=1.2\linewidth,
                           height=6.8cm,
                           keepaspectratio]{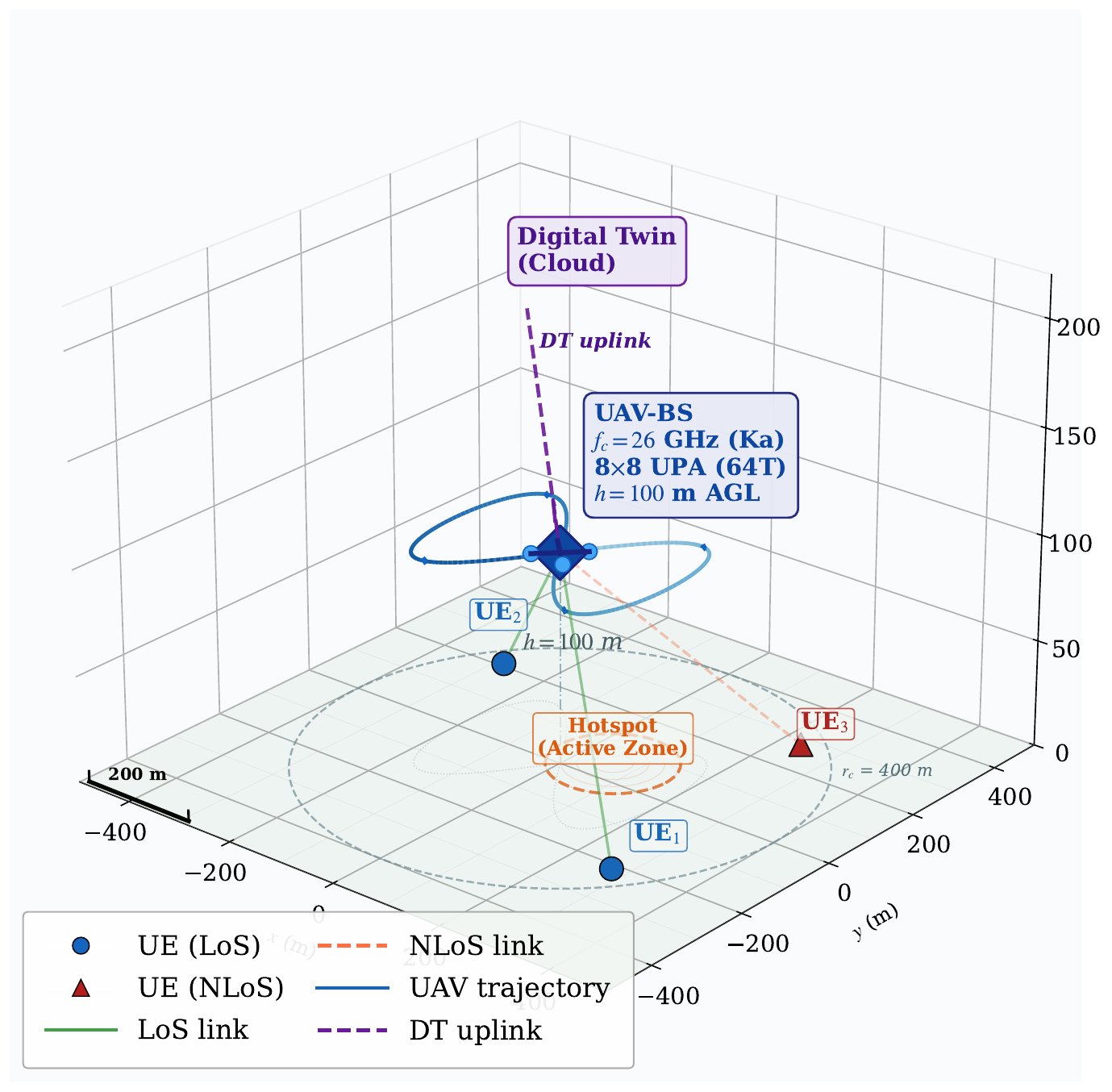}}
    \caption{UAV-BS 6G NTN scenario}
    \label{fig:UAV_DT}
\end{subfigure}
\hfill
\begin{subfigure}[b]{0.48\textwidth}
    \centering
    \fbox{\includegraphics[width=1.2\linewidth,
                           height=6.8cm,
                           keepaspectratio]{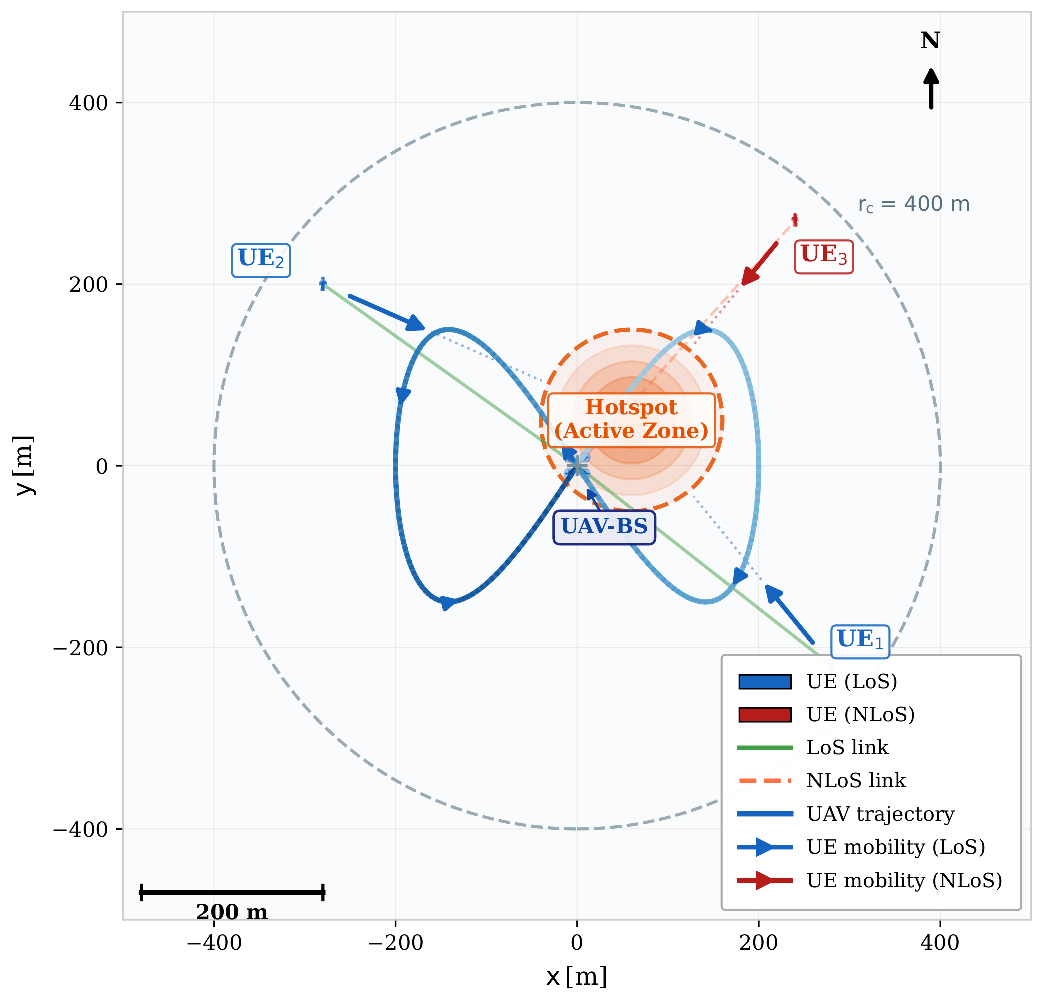}}
    \caption{Top-view: UAV trajectory and user hotspot}
    \label{fig:topview}
\end{subfigure}

\vspace{4mm}

\begin{subfigure}[b]{0.48\textwidth}
    \centering
    \fbox{\includegraphics[width=0.96\linewidth,
                           height=4.8cm,
                           keepaspectratio]{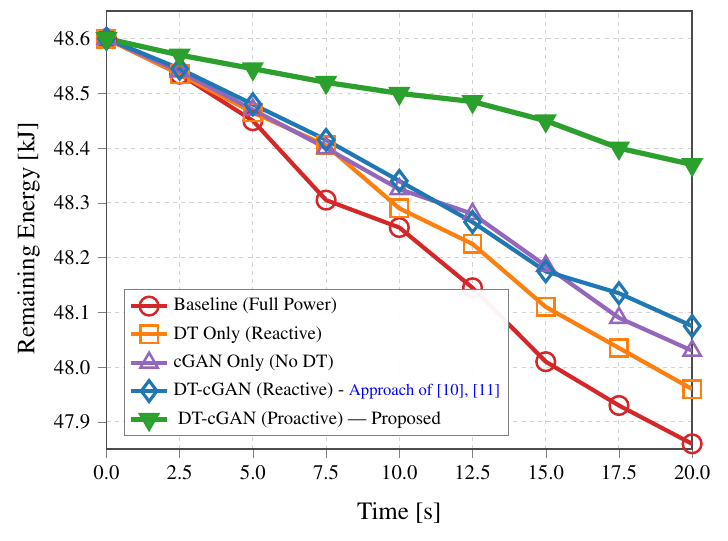}}
    \caption{Remaining battery energy over time}
    \label{fig:batt_energy}
\end{subfigure}
\hfill
\begin{subfigure}[b]{0.48\textwidth}
    \centering
    \fbox{\includegraphics[width=0.96\linewidth,
                           height=4.8cm,
                           keepaspectratio]{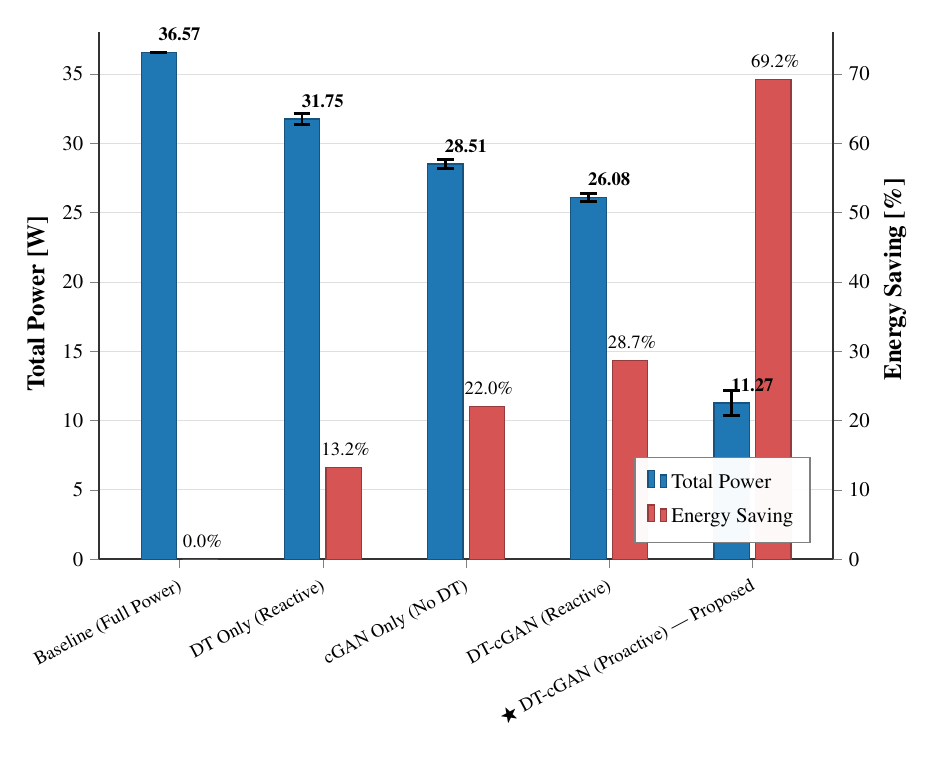}}
    \caption{Total power and energy-saving comparison}
    \label{fig:total_power}
\end{subfigure}

\vspace{3mm}

\caption{Energy-efficiency evaluation of the proposed proactive 
predictive DT framework in a UAV-assisted NTN deployment 
under 3GPP TR~28.915/38.821 at 100~m altitude above ground 
level (AGL).
(a)~3D UAV-BS NTN scenario.
(b)~Top-view of the UAV-BS flight trajectory.
(c)~Remaining UAV battery energy over the mission duration for 
the proposed and all benchmark schemes.
(d)~Total power consumption and percentage energy savings 
achieved by each compared framework.}

\label{fig:energyeval}

\end{figure*}

\section{Illustrative Deployment in UAV-Assisted Non-Terrestrial Networks}
\label{sec:deployment}

The proposed predictive decision engine is a general system-level framework for AI-native wireless networks. Rather than validating every application individually, this section demonstrates its practical realization through a representative deployment in a UAV-assisted non-terrestrial network (NTN), chosen because it combines rapid topology variations, dynamic wireless channels, user mobility, and stringent onboard energy constraints within a single scenario. These characteristics stress the framework's three fundamental components, namely continuous GenAI--DT synchronization, predictive intelligence, and closed-loop wireless control. Although this case study focuses on UAV energy management using a cGAN as the illustrative predictor, the underlying architectural principles apply directly to other AI-native 6G scenarios, including ISAC, cell-free massive MIMO, semantic communications, smart manufacturing, and integrated terrestrial--non-terrestrial networks, regardless of which predictive model is used.

\subsection{Illustrative Deployment Scenario}

We consider a 3GPP-compliant low-altitude NTN deployment at a Ka-band carrier frequency of 26~GHz, following 3GPP TR~28.915 and 38.821~\cite{3gpp28915, 3gpp38821}. A UAV-mounted base station with an $8\times8$ uniform planar array (UPA) covers a $400$~m service area while flying at a nominal altitude of $100$~m. Ground users follow a clustered random waypoint mobility model at $1$~m/s pedestrian speed. Figures~\ref{fig:UAV_DT} and~\ref{fig:topview} show the three-dimensional deployment and top-view geometry. The physical network streams heterogeneous observations, including CSI, user mobility, traffic evolution, propagation conditions, and battery state-of-charge (SoC), to the synchronized DT, which fuses them into a continuously evolving twin-state. Consistent with Section~\ref{sec:framework}, the predictive engine reasons over this state to anticipate future channel quality, congestion, interference, and energy demand before control decisions are generated.

For this deployment, a lightweight cGAN serves as the predictive engine, chosen for its low inference latency and suitability for real-time wireless control. The generator is conditioned on synchronized DT states, including UAV position, mobility features, estimated channel quality, traffic demand, and battery SoC, while the discriminator keeps predicted network evolution statistically consistent with realistic wireless behavior. During online operation, predicted states are forwarded to the decision layer, where transmission power, beamforming, scheduling, and energy-aware resource allocation are continuously adapted before degradation occurs: when excess signal-to-interference-plus-noise ratio (SINR) margin is predicted, transmission power is proactively reduced to conserve UAV battery energy, and when channel degradation or congestion is anticipated, resources are restored before quality-of-service (QoS) violations occur.

\subsection{Demonstrating Predictive Closed-Loop Wireless Control}

This deployment demonstrates how the predictive decision engine translates synchronized awareness into proactive wireless control. Rather than reacting to instantaneous channel conditions, the framework continuously predicts future network evolution and adjusts transmission policies before performance deteriorates, which matters most in UAV-assisted NTN deployments where battery capacity, mobility, and rapidly changing propagation jointly determine performance.

Figure~\ref{fig:batt_energy} shows the temporal evolution of transmission power and cumulative energy consumption. Unlike the reactive benchmark, which continuously transmits at conservative power levels to accommodate unexpected channel degradation, the proposed framework exploits synchronized DT information and predictive inference to identify upcoming favorable channel conditions and proactively reduce power while maintaining QoS, restoring power before service quality is affected when degradation or rising traffic demand is anticipated.

Figure~\ref{fig:total_power} summarizes the overall energy benefits. The reactive maximum ratio transmission (MRT) baseline shows the highest power consumption, since decisions rely only on the current state. DT synchronization alone gives moderate savings through improved environmental awareness, while the cGAN-only configuration benefits from prediction but lacks continuous synchronization, combining both yields further improvement. This reactive DT–cGAN configuration is architecturally representative of the Stage~II category in Table~\ref{tab:related},  such as, \cite{singh_wind_2025, cao_decentralized_2026}, where prediction remains coupled to the synchronized twin but decision-making is executed as a separate, reactive step; the proposed proactive framework extends this by fusing decision-making directly into the closed loop, consistent with the Stage III redefinition. The proposed framework achieves the largest reduction, lowering average transmission power from approximately 36.57~W to 11.27~W, an energy reduction of nearly 69.2\% relative to the reactive baseline. These improvements should be read as validation of the architectural redefinition described in Section~\ref{sec:framework} rather than of the predictive model alone, since the exact numerical gains will depend on synchronization accuracy, prediction quality, mobility, traffic dynamics, and deployment density, but the underlying architectural principle remains unchanged.

\subsection{Complexity and Sensitivity Considerations}
\label{sec:complexity}

Table~\ref{tab:complexity} reports representative complexity figures for a lightweight cGAN matching the framework above, implemented as a compact residual generator (three residual blocks, 128 hidden units) and a comparably small discriminator, profiled on a general-purpose CPU. The generator, the only component required at deployment since the discriminator is used solely during offline training, contains approximately 106K parameters (0.43~MB in single precision, 106.9K multiply-accumulate operations (MACs) per inference) and completes a single-instance prediction in under 0.1~ms, well within the latency budget of a UAV control loop operating on the order of tens of milliseconds. Batching inference across multiple simultaneously tracked entities amortizes this cost further, reducing per-instance latency to roughly 3~$\mu$s at a batch of 100, indicating that one edge-deployed instance can plausibly support many concurrently synchronized twins. These figures characterize a representative lightweight instantiation rather than an exhaustive profiling of every possible backbone comprising of heavier predictors, such as diffusion models or large multimodal foundation models, which would trade this latency and memory footprint for richer reasoning capacity, consistent with the model-agnostic design of Section~\ref{sec:framework}.

\begin{table}[t]
\caption{Representative Complexity of the Illustrative cGAN Predictor.}
\label{tab:complexity}
\definecolor{dtrow}{RGB}{235,244,255}
\definecolor{sysrow}{RGB}{255,248,230}
\definecolor{proprow}{RGB}{228,244,228}
\definecolor{headercolor}{RGB}{30,80,160}
\centering
\scriptsize
\renewcommand{\arraystretch}{1.15}
\begin{tabular}{@{}lccc@{}}
\toprule
\rowcolor{headercolor}
\textcolor{white}{\textbf{Metric}} & \textcolor{white}{\textbf{Generator}} & \textcolor{white}{\textbf{Discriminator}} \\
\midrule
\rowcolor{dtrow}
Parameters & 106,384 & 38,401 \\
\rowcolor{dtrow}
MACs / inference & 106.9K & 38.8K \\
\rowcolor{dtrow}
Model size (fp32) & 0.43~MB & 0.15~MB \\
\rowcolor{dtrow}
Latency, batch = 1 (CPU) & 0.09~ms & 0.04~ms \\
\rowcolor{dtrow}
Latency, batch = 100 (CPU) & 0.003~ms/instance & -- \\
\bottomrule
\end{tabular}
\end{table}

A full sensitivity study across prediction error, synchronization delay, mobility models, and user density is left for future work, but the framework's design suggests illustrative trends. Since control decisions depend on the predicted margin, moderate noise on the predicted state should erode but not eliminate energy savings until it approaches the operating margin, at which point performance degrades gracefully toward the reactive baseline. Synchronization delay acts similarly, for example, a delay of $\delta$ shortens the usable prediction horizon by $\delta$, so performance should decline gradually as $\delta$ approaches the horizon length rather than fail sharply. Denser deployments and richer mobility patterns raise synchronization and inference overhead roughly with the number of tracked entities, an effect the batching results above suggest can be largely absorbed through shared inference. A systematic sensitivity study is a natural next step once the framework is validated on physical NTN testbeds.

\section{Conclusions and Future Research Directions}
This article redefines the DT from a synchronized replica into a predictive decision engine within a unified, closed-loop, AI-native system, independent of any specific underlying learning model. Use-case study included a representative UAV-assisted NTN deployment, using a lightweight cGAN as one illustrative predictor. It demonstrated the resulting energy-efficiency gains alongside a lightweight complexity and sensitivity characterization. Although the case study focused on UAV communications, the framework is application-independent and extends naturally to other AI-native 6G scenarios, and maps onto the network DT work underway in 3GPP Releases~19 and~20. Future work should pursue reasoning-capable foundation models, multimodal LLMs, autonomous AI agents, distributed and cooperative DTs, systematic sensitivity and complexity benchmarking, along with field validation of DT fidelity against physical measurements. These will be key steps toward fully autonomous AI-native 6G networks capable of continuously sensing, predicting, deciding, acting, and learning across large-scale cyber--physical wireless environments.

\bibliographystyle{IEEEtran}
\bibliography{Ref}

\end{document}